\documentclass[12pt,a4paper]{article}
\usepackage{amssymb}
\newcommand{\R}{\mathbb{R}}
\newcommand{\Z}{\mathbb{Z}}
\newcommand{\N}{\mathbb{N}}
\newcommand{\epn}{\epsilon}
\newcommand{\p}{\partial}
\newcommand{\nn}{\nonumber\\}
\newcommand{\gIIB}{g_s^{\rm{IIB}}}
\newcommand{\gsb}{g_b}
\newcommand{\cpq}{c_{pq}}
\newcommand{\Opq}{O_{(p,q)}}
\newcommand{\hatg}{\hat{\gamma}}
\newcommand{\hmu}{\hat{\mu}}
\newcommand{\hnu}{\hat{\nu}}
\newcommand{\hp}{\hat{p}}
\newcommand{\hq}{\hat{q}}
\newcommand{\hA}{\hat{A}}
\newcommand{\hB}{\hat{B}}
\newcommand{\hC}{\hat{C}}
\newcommand{\hM}{\hat{M}}
\newcommand{\hN}{\hat{N}}
\newcommand{\hP}{\hat{P}}
\newcommand{\hG}{G}

\newcommand{\ga}{g}
\newcommand{\tg}{\tilde{g}}
\newcommand{\tG}{\tilde{G}}
\newcommand{\tX}{\tilde{X}}
\newcommand{\BNS}{B^{(1)}}
\newcommand{\BR}{B^{(2)}}
\newcommand{\Bpq}{B^{(pq)}}

\title{($p,q$)-string in the wrapped supermembrane on 2-torus\\
{}-- {\sl\large A classical analysis of the bosonic sector} --}
\author{
{\sc Hiroyuki Okagawa}\footnote{email:
	okagawa@eken.phys.nagoya-u.ac.jp},~
{\sc Shozo Uehara}\footnote{e-mail:
	uehara@eken.phys.nagoya-u.ac.jp}~ and
{\sc Satoshi Yamada}\footnote{e-mail:
	yamada@eken.phys.nagoya-u.ac.jp}\vspace{4mm}\\
{\sl Department of Physics, Nagoya University}\\
{\sl Chikusa-ku, Nagoya 464-8602, Japan}}
\date{}
\makeatletter

 \@addtoreset{equation}{section}
\renewcommand{\thefigure}{\@arabic\c@figure}
\makeatother
\begin{document}
\maketitle
\vspace{-80mm}
\begin{flushright}
	DPNU-06-01\\
	hep-th/0603203\\
	March 2006
\end{flushright}
\vspace{57mm}

\begin{abstract}
We consider a wrapped supermembrane on $\R^9\times T^2$.
We examine a double dimensional reduction to deduce a $(p,q)$-string
in type IIB superstring theory from the wrapped supermembrane.
In particular, directly from the wrapped supermembrane action, we
explicitly derive the action of a string which carries the RR 2-form
charge as well as the NSNS 2-form charge, and the tension of the
string agrees with the $(p,q)$-string tension.
\end{abstract}

\newpage
%%%%%%%%%%%%%%%%%%%%%%%%%%%%%%%%%%%
\section{Introduction}
M-theory includes the supermembrane in eleven dimensions \cite{BST}
which is expected to play an important role to understand the
fundamental degrees of freedom in M-theory.
Actually, it was shown that the wrapped supermembrane on
$\R^{10}\times S^1$ is related to type IIA superstring on $\R^{10}$
by means of the double dimensional reduction \cite{DHIS}.
On the other hand, type IIB superstring is related to type IIA
superstring via T-duality, or type IIA superstring on $\R^9\times S^1$
leads to type IIB superstring on $\R^{10}$ in the shrinking limit of
the $S^1$ radius. Hence, type IIB superstring in 10 dimensions is to
be deduced from supermembrane on a vanishing 2-torus.

Schwarz showed an $SL(2,\Z)$ family of string solutions of type IIB
supergravity \cite{Sch}. The $(p,q)$-string \cite{Sch,W} is considered
to be the bound state of fundamental strings (F-strings) and D1-branes
(D-strings) in type IIB superstring theory.
Furthermore, it was pointed out that the supermembrane which is
wrapping $p$-times around one of two compact directions and $q$-times
around the other direction gives a $(p,q)$-string. However, it has not
been derived directly from the supermembrane action.
In this paper we consider shrinking the 2-torus to approach type IIB
superstring theory.
Actually we deduce $(p,q)$-strings in type IIB superstring theory from
the wrapped supermembrane on $\R^9\times T^2$ in the shrinking limit
of the 2-torus.

The plan of this paper is as follows. In the next section, we consider
the supermembrane on $\R^9\times T^2$.
We shall carefully rewrite the eleven-dimensional supergravity
background fields to the nine dimensional ones and consider the double
dimensional reduction along an oblique direction of $T^2$.
In section \ref{S:C}, we consider the T-dual of the derived string
action along another compact direction of the 2-torus to deduce a
string action with the $(p,q)$-string tension.
We shall see that the string carries $p$-times the unit NSNS 2-form
charge and $q$-times the unit RR 2-form charge as well, which
indicates that the deduced string is, in fact, a $(p,q)$-string in
type IIB superstring theory.
The final section contains some discussion.

%%%%%%%%%%%%%%%%%%%%%%%%%%%%%%%%%%%%%%%%%%%%%%%%
\section{Double dimensional reduction}\label{S:F}
%%%%%%%%%%%%%%%%%%%%%%%%%%%%%%%%%%%%%%%%%%%%%%%%
The supermembrane in a eleven-dimensional supergravity background is
given by \cite{BST}
\begin{eqnarray}
  S&=&T\int\!d\tau\!\int_0^{2\pi}\!d\sigma d\rho
   \Biggl[\frac{1}{2}\sqrt{-\hatg}\,\hatg^{\alpha\beta}\,
	E_\alpha^{~A} E_\beta^{~B}\,\eta_{AB}\nn
  &&\hspace{20ex}-\frac{1}{2}\sqrt{-\hatg}
    -\frac{1}{3!}\,\epn^{\alpha\beta\gamma}\,E_\alpha^{~\hA}
    E_\beta^{~\hB} E_\gamma^{~\hC}\hA_{\hC\hB\hA}\Biggr]
	\label{eq:SMac},
\end{eqnarray}
where $T$ is the tension of the supermembrane, $\hatg_{\alpha\beta}\
(\alpha,\beta=0,1,2)$ is the worldvolume metric,
$\hatg=\det\hatg_{\alpha\beta}$, and the target space is a
supermanifold with the superspace coordinates
$Z^{\hM}=(X^M,\theta^{m})~(M=0,\cdots,10,~m=1,\cdots,32)$.
Furthermore, $\hA_{\hM\hN\hP}(Z)$ is the super three-form and
$E_\alpha^{~\hA}\equiv(\p_\alpha Z^{\hM})\,E_{\hM}^{~\hA}$ where
$E_{\hM}^{~\hA}$ is the supervielbein and $\hA=(A,a)$ is the tangent
space index.
We shall consider a dimensional reduction and in order that one can
see the procedure easily we focus on the bosonic degrees of freedom
hereafter.
The bosonic background fields are included in the superfields as
\begin{equation}
    E_M^{~A}(Z)\bigg|_{\rm{fermions}=0}= e_M^{~A}(X)\,,\quad
    \hA_{MNP}(Z)\bigg|_{\rm{fermions}=0}= A_{MNP}(X)\,.
\end{equation}
Then the action (\ref{eq:SMac}) is reduced to\footnote{The
mass dimensions of the worldvolume parameters ($\tau,\sigma$,
$\rho$) and the eleven-dimensional background fields ($G_{MN}$,
$A_{MNP}$) are $0$. And the mass dimension of worldvolume metric
$\hatg_{\alpha\beta}$ is $-2$.}
\begin{eqnarray}
  S&=&T\int\!d\tau\!\int_0^{2\pi}\!d\sigma d\rho
   \Biggl[\frac{1}{2}\sqrt{-\hatg}\,\hatg^{\alpha\beta}\,
	\p_\alpha X^M\p_\beta X^N\,\hG_{MN}(X)\nn
  &&\hspace{20ex}-\frac{1}{2}\sqrt{-\hatg}
    +\frac{1}{3!}\,\epn^{\alpha\beta\gamma}\,\p_\alpha X^M
    \p_\beta X^N\p_\gamma X^P A_{MNP}(X)\Biggr]\label{eq:Mac}.
\end{eqnarray}
Note that variation w.r.t.\ $\hatg_{\alpha\beta}$ yields the induced
metric,
\begin{equation}
  \hat\gamma_{\alpha\beta}=\p_\alpha X^M \p_\beta X^N\,\hG_{MN}(X)\,,
\end{equation}
and plugging it back into the original action leads to the Nambu-Goto
form
\begin{equation}
  S=T\int\!d\tau\!\int_0^{2\pi}\!d\sigma d\rho
   \Biggl[\sqrt{-\hatg}
    +\frac{1}{3!}\,\epn^{\alpha\beta\gamma}\,\p_\alpha X^M
    \p_\beta X^N\p_\gamma X^P\, A_{MNP}(X)\Biggr]\,.\label{eq:Mac1}
\end{equation}
Actually, we consider a wrapped supermembrane action
on $\R^9\times T^2$. We shall take the shrinking limit
of the 2-torus and deduce the $(p,q)$-string action directly from the
action.
We take the 10th and 9th directions to compactify on $T^2$,
whose radii are $L_1$ and $L_2$, respectively.
In taking the shrinking volume limit of the 2-torus, we keep the ratio
of the radii finite,
\begin{equation}
  \gsb\equiv \frac{L_1}{L_2}~\mbox{: finite.}\quad (L_1,L_2\to0)
\end{equation}

Now we consider two cycles on $T^2$ characterized by two sets of
co-prime integers ($p,q$) and ($r,s$). We impose the following
condition on the two sets of co-prime integers in order that one can
adjust
the spacesheet coordinates to the target
space,\footnote{Geometrically, the first condition in eq.(2.5)
indicates the orthogonality of the two vectors ($p,q$) and ($r,s$) and
the
second condition means that the area defined by the two vectors is
non-zero.}
\begin{equation}
  pr+qs=0,\quad ps-qr\ne0,\quad p,q,r,s \in\Z\,.
\end{equation}
Considering the line-element in $(X^9,X^{10})$ surface,
\begin{eqnarray}
 && \hG_{99}(dX^9)^2 + 2\hG_{910}dX^9 dX^{10}
	+\hG_{1010}(dX^{10})^2\nn
 &&\qquad=\Biggl(\hG_{99}
    -\frac{(\hG_{910})^2}{\hG_{1010}}\Biggr)\,(dX^9)^2
	+\hG_{1010}\,\Biggl(dX^{10}
		+\frac{\hG_{910}}{\hG_{1010}}\,dX^9\Biggr)^2\,,
\end{eqnarray}
we shall represent the wrapping of the supermembrane as
\begin{eqnarray}
 \sqrt{\hG_{1010}}\,X^{10}(\tau,\sigma,\rho+2\pi)&=&2\pi w_1 L_1 p
    +\sqrt{\hG_{1010}}\,X^{10}(\tau,\sigma,\rho)\,,\label{eq:pq1p}\\
	\sqrt{\hG_{99}-\frac{(\hG_{910})^2}{\hG_{1010}}}\,
	X^{9}(\tau,\sigma,\rho+2\pi)
  &=&2\pi w_1 L_2 q +\sqrt{\hG_{99}-\frac{(\hG_{910})^2}{\hG_{1010}}}
	\,X^{9}(\tau,\sigma,\rho)\,,\label{eq:pq1q}\\
 \sqrt{\hG_{1010}}\,X^{10}(\tau,\sigma+2\pi,\rho)&=& 2\pi w_2 L_1 r
    +\sqrt{\hG_{1010}}\,X^{10}(\tau,\sigma,\rho)\,,\label{eq:pq2r}\\
	\sqrt{\hG_{99}-\frac{(\hG_{910})^2}{\hG_{1010}}}\,
	X^{9}(\tau,\sigma+2\pi,\rho)
  &=& 2\pi w_2 L_2 s +\sqrt{\hG_{99}-\frac{(\hG_{910})^2}{\hG_{1010}}}
	\,X^{9}(\tau,\sigma,\rho)\,,\label{eq:pq2s}
\end{eqnarray}
or
\begin{eqnarray}
 X^{10}(\xi^\alpha)&=&
    \frac{L_1\,(w_1p\rho+w_2r\sigma)}{\sqrt{\hG_{1010}}}
	+Y^1(\xi^\alpha)\,,\label{eq:pq1}\\
 X^9(\xi^\alpha)&=&\frac{L_2\,(w_1q\rho+w_2s\sigma)}{\sqrt{\hG_{99}
	-\frac{(\hG_{910})^2}{\hG_{1010}}}}
	+Y^2(\xi^\alpha)\,,\label{eq:pq2}
\end{eqnarray}
where $\xi^{\alpha}=\tau,\sigma,\rho$ and
\begin{equation}
    w_n\in\N\,.\quad(n=1,2)\label{eq:WN}
\end{equation}
Note that $w_n$ can be negative, or $w_n\in\Z\backslash\{0\}$,
however, one can flip the signs of $p,q\to -p,-q$ (for $w_1$) and
$r,s\to -r,-s$ (for $w_2$) to have eq.(\ref{eq:WN}) without loss of
generality.
The above equations indicate that the supermembrane is wrapping
$w_1p$-times around one of the two compact directions ($X^{10}$)
and $w_1 q$-times around the other direction ($X^{9}$) if one advances
by $2\pi$ along the $\rho$-direction on the worldsheet. Thus, this
wrapped supermembrane is expected to give $(p,q)$-strings \cite{Sch}.
Actually, we shall see that the $(p,q)$-string comes out through the
double dimensional reduction and T-duality in the next section.
 
Now that we shall adopt the double dimensional reduction technique
\cite{DHIS}. However, we should be careful to deduce
$(p,q)$-strings. First we determine the spacetime direction to align
with one of the worldvolume coordinate, or we fix the gauge.
We define $X^y$ and $X^z$ by an SO(2) rotation of the target space,
\begin{equation}
  \left(\begin{array}{@{\,}c@{\,}} X^z \\ X^y \end{array} \right)=\Opq
    \left(\begin{array}{@{\,}c@{\,}}X^{10} \\ X^9 \end{array}\right),
	\label{eq:so2rot}
\end{equation}
where
\begin{equation}
  \Opq =\frac{1}{\cpq}\left(\begin{array}{@{\,}cc@{\,}}
        p & q\\  -q & p \end{array}\right)
	\equiv \left(\begin{array}{@{\,}cc@{\,}}
        \hp & \hq\\  -\hq & \hp \end{array}\right)\in SO(2)\,,
 \quad\cpq\equiv \sqrt{p^2 +q^2}\,. \label{eq:so2mx}
\end{equation}
By using the relations between the eleven-dimensional supergravity and
nine-dimensional (or $S^1$-compactified) type IIB fields
\cite{BHO,MO}, we have
\begin{equation}
  \sqrt{\frac{\hG_{99}-\frac{(\hG_{910})^2}{\hG_{1010}}}{\hG_{1010}}}
  =  e^{-\varphi}=\gsb^{-1}=\frac{L_2}{L_1}\,,
\end{equation}
where $\varphi$ is a type IIB dilaton background. Then we have
\begin{eqnarray}
  X^z&=& \frac{L_1\,w_1\cpq\,\rho}{%
    \sqrt{\hG_{1010}}} +\hp\,Y^1(\xi^\alpha)+\hq\,Y^2(\xi^\alpha)\,,\\
  X^y&=& \frac{L_2\,(ps-qr)w_2\,\sigma}{\cpq\sqrt{\hG_{99}
	-\frac{(\hG_{910})^2}{\hG_{1010}}}}
	-\hq\,Y^1(\xi^\alpha)+\hp\,Y^2(\xi^\alpha)\,.
\end{eqnarray}

A suitable choice of the target-space metric is
($M,N=0,1,\cdots,8,9,10$)
\begin{eqnarray}
 \hG_{MN} &=& \left(\begin{array}{@{\,}cc@{\,}}
	\frac{1}{\sqrt{\hG_{1010}}}\,\ga_{\hmu\hnu}
	+\frac{1}{\hG_{1010}}\hG_{\hmu10}\hG_{\hnu10}&
		 \hG_{\hmu10} \\[10pt]
	\hG_{\hnu10} & \hG_{1010} \end{array}\right)\nn
  &=& \left(\begin{array}{@{\,}ccc@{\,}}
	\frac{1}{\sqrt{\hG_{1010}}}\,\ga_{\mu\nu}
	+\frac{1}{\hG_{1010}}\hG_{\mu10}\hG_{\nu10}&
	\frac{1}{\sqrt{\hG_{1010}}}\,\ga_{\mu9}
	+\frac{1}{\hG_{1010}}\hG_{\mu10}\hG_{910}&\hG_{\mu10} \\[10pt]
	\frac{1}{\sqrt{\hG_{1010}}}\,\ga_{9\nu}
	+\frac{1}{\hG_{1010}}\hG_{910}\hG_{\nu10}&
	\frac{1}{\sqrt{\hG_{1010}}}\,\ga_{99}
	+\frac{1}{\hG_{1010}}\hG_{910}\hG_{910}& \hG_{910} \\[10pt]
	\hG_{\nu10}&\hG_{910} & \hG_{1010}
			\end{array}\right),\quad
\end{eqnarray}
where $\hmu,\hnu=0,1,\cdots,8,9$ and $\mu,\nu=0,1,\cdots,8$.
On the other hand, due to eq.(\ref{eq:so2rot}) we may also write
($U,V=0,1,\cdots,8,y,z$)
\begin{eqnarray}
 \tG_{UV} &=& \hG_{MN}\,\frac{\p X^M}{\p X^U}\,
	\frac{\p X^N}{\p X^V}\nn
  &=& \left(\begin{array}{@{\,}ccc@{\,}}
	\frac{1}{\sqrt{\tG_{zz}}}\,\tg_{\mu\nu}
	+\frac{1}{\tG_{zz}}\tG_{\mu z}\tG_{\nu z}&
	\frac{1}{\sqrt{\tG_{zz}}}\,\tg_{\mu y}
	+\frac{1}{\tG_{zz}}\tG_{\mu z}\tG_{yz}& \tG_{\mu z} \\[10pt]
	\frac{1}{\sqrt{\tG_{zz}}}\,\tg_{y\nu}
	+\frac{1}{\tG_{zz}}\tG_{yz}\tG_{\nu z}&
	\frac{1}{\sqrt{\tG_{zz}}}\,\tg_{yy}
	+\frac{1}{\tG_{zz}}\tG_{yz}\tG_{yz}& \tG_{yz} \\[10pt]
	\tG_{\nu z}&\tG_{yz} & \tG_{zz}	\end{array}\right)\,,
\end{eqnarray}
and hence we have
\begin{eqnarray}
 \tG_{zz}&=&\hq^2\,\hG_{99}+2\hp\hq\,\hG_{910}
		+\hp^2\,\hG_{1010}\,, \\
 \tG_{yy}&=&\hp^2\,\hG_{99}-2\hp\hq\,\hG_{910}
		+\hq^2\,\hG_{1010}\,,\\
 \tG_{yz}&=&\hp\hq\,\hG_{99}+(\hp^2-\hq^2)\,\hG_{910}
		-\hp\hq\,\hG_{1010}\,.
\end{eqnarray}
Now we shall make a (partial) gauge choice of (cf. Ref.\cite{DHIS})
\begin{equation}
  X^z=\frac{L_1 w_1\cpq}{\sqrt{\hG_{1010}}}\,\rho\,
	\equiv C\rho,\label{eq:DDR0}
\end{equation}
or the $z$-direction is aligned with one of the space direction $\rho$
of the worldvolume. Then the dimensional reduction is achieved by
imposing the following conditions on the membrane-coordinates and the
background fields,
\begin{equation}
  \p_\rho X^y = \p_\rho X^\mu=0\,,\label{eq:DDR1}
\end{equation}
and
\begin{equation}
  \p_z \hG_{MN} = \p_z A_{MNP} =0\,.\label{eq:DDR2}
\end{equation}
Thus the induced metric on the worldvolume is given by \cite{DHIS}
\begin{equation}
 \hat\gamma_{\alpha\beta}=\p_\alpha X^M \p_\beta X^N\,\hG_{MN}(X)
  =\Phi^{-2/3}\left(\begin{array}{@{\,}cc@{\,}}
	\gamma_{ij}+ \Phi^2 A_iA_j & \Phi^2 A_i \\[10pt]
	\Phi^2A_j & \Phi^2 \end{array}\right)\,,
\end{equation}
where $i,j=0,1$ and
\begin{eqnarray}
 \Phi^{4/3} &=& C^2\,\tG_{zz}\,,\\
 \Phi^{4/3}A_i
	&=& C(\p_iX^{\mu} \,\tG_{\mu z}+\p_iX^y \,\tG_{yz})\,,\\
 \gamma_{ij}
    &=&C\,(\p_iX^\mu\p_jX^\nu\,\tg_{\mu\nu}+2\p_{\{i}X^\mu \p_{j\}}X^y
	\tg_{\mu y}+\p_iX^y \p_jX^y \tg_{yy})\,.
\end{eqnarray}
Note that
\begin{equation}
  \det\hat\gamma_{\alpha\beta}=\det\gamma_{ij}\,,
\end{equation}
which is to be used in calculating the first term in
eq.(\ref{eq:Mac1}). From eqs.(\ref{eq:so2rot}), (\ref{eq:DDR0}) and
(\ref{eq:DDR1}), we have
\begin{eqnarray}
 &&\hspace{-5ex}\epn^{\alpha\beta\gamma}\,\p_\alpha X^M
    \p_\beta X^N\p_\gamma X^P A_{MNP}\nn
 &=&3C\{\,\epn^{ij}\,\p_i X^\mu \p_j X^\nu
	(\hq A_{\mu\nu9}+\hp A_{\mu\nu10})
 	+2\,\epn^{ij}\,\p_i X^\mu\p_j X^y A_{\mu910}\}\,.
\end{eqnarray}
Thus, by the double dimensional reduction of
eqs.(\ref{eq:DDR0})-(\ref{eq:DDR2}), the supermembrane action
(\ref{eq:Mac}), or (\ref{eq:Mac1}), is reduced to the following
equivalent one,
\begin{eqnarray}
 S_{ddr}&=&\frac{2\pi T}{2}\int\!d\tau\!\int_0^{2\pi}
    \hspace{-1ex}d\sigma\,C\Biggl[\sqrt{-\tilde\gamma}\,
      \tilde\gamma^{ij}\,(\p_iX^\mu\p_jX^\nu\,\tg_{\mu\nu}
      +2\p_iX^\mu\p_j X^y\tg_{\mu y}+\p_iX^y \p_jX^y\tg_{yy})\nn
  &&\hspace{10ex}+\{\epn^{ij}\,\p_i X^\mu \p_j X^\nu
    (\hq A_{\mu\nu9}+\hp A_{\mu\nu10})
    +2\epn^{ij}\,\p_i X^\mu\p_j X^y A_{\mu910}\}\Biggr],\label{eq:Macr}
\end{eqnarray}
where the first term on the r.h.s.\ has been rewritten in Polyakov
form by introducing the worldsheet metric $\tilde\gamma_{ij}$ instead
of Nambu-Goto form. As is pointed out in \cite{DHIS}, this action
(\ref{eq:Macr}) has conformal invariance.

%%%%%%%%%%%%%%%%%%%%%%%%%%%%%%%%%%%%%%%%%%%%%%%%
\section{\boldmath$(p,q)$-string from wrapped
supermembrane}\label{S:C}
%%%%%%%%%%%%%%%%%%%%%%%%%%%%%%%%%%%%%%%%%%%%%%%%
In this section, we derive the $(p,q)$-string action from the
reduced supermembrane action in eq.(\ref{eq:Macr}).
We shall take T-dual along the other compactified $X^y$-direction
(cf. eq.(\ref{eq:so2rot})).
Introducing a variable $\tX^y$, which is seen to be dual to the other
compactified $X^y$-direction, eq.(\ref{eq:Macr}) can be rewritten by
\begin{eqnarray}
 S_{ddr}&=&\frac{2\pi T}{2}\int\!d\tau\!\int_0^{2\pi}
    \hspace{-1ex}d\sigma\,C\Biggl[\sqrt{-\tilde\gamma}\,
      \tilde\gamma^{ij}\,(\p_iX^\mu\p_jX^\nu\,\tg_{\mu\nu}
      +2\p_{i}X^\mu Y_j\,\tg_{\mu y} +Y_i Y_j\,\tg_{yy})\nn
  &&\hspace{1ex}+\{\epn^{ij}\,\p_i X^\mu \p_j X^\nu
    (\hq A_{\mu\nu9}+\hp A_{\mu\nu10})
	+2\epn^{ij}\,\p_i X^\mu Y_j A_{\mu910}
	+2\epn^{ij}\tX^y\p_iY_j\}\Biggr]\,,~\label{eq:d-Macr}
\end{eqnarray}
since the variation w.r.t.\ $\tX^y$ leads to $\epn^{ij}\p_iY_j=0$ or
$Y_j=\p_j X^y$ and hence eq.(\ref{eq:Macr}) can be
reproduced.\footnote{We assume that the background fields are
independent of $\tX^y$ in eq.(\ref{eq:d-Macr}).}
On the other hand, assuming that all the fields are independent of
$Y_j$ (or $X^y$), the variation w.r.t.\ $Y_i$ leads to
\begin{equation}
   Y_i = -\frac{\tilde\gamma_{ij}\epn^{jk}}{\sqrt{-\tilde\gamma}
	\,\tg_{yy}}\,\left(\p_k\tX^y-A_{\mu910}\p_kX^\mu\right)
	-\frac{\tg_{y\mu}}{\tg_{yy}}\,\p_i X^\mu,
\end{equation}
and hence we have
\begin{eqnarray}
 S_{ddr}&=&\frac{2\pi T}{2}\int\!d\tau\!\int_0^{2\pi}
    \hspace{-1ex}d\sigma\,C\Biggl[\sqrt{-\tilde\gamma}\,
      \tilde\gamma^{ij}\,\Biggl\{\p_iX^\mu\p_jX^\nu\,\Bigl(\tg_{\mu\nu}
    -\frac{\tg_{y\mu}\tg_{y\nu}-A_{\mu910}A_{\nu910}}{\tg_{yy}}\Bigr)\nn
  &&\quad-2\p_i X^\mu\p_j \tX^y\,\frac{A_{\mu910}}{\tg_{yy}}
	+\p_i\tX^y \p_j\tX^y\,\frac{1}{\tg_{yy}}\Biggr\}
	+\epn^{ij}\,\p_i X^\mu \p_j X^\nu
	\Bigl(\hq A_{\mu\nu9}+\hp A_{\mu\nu10}\nn
  &&\quad-\frac{2A_{\mu910}\tg_{\nu y}}{\tg_{yy}}\Bigr)
	+2\,\epn^{ij}\,\p_i\tX^y\p_jX^\mu\,
	\frac{\tg_{y\mu}}{\tg_{yy}}\Biggr]\,.\label{eq:ddrac}
\end{eqnarray}

Now that we consider T-dual for the background fields in
eq.(\ref{eq:Macr}) (or eq.(\ref{eq:ddrac})).
Since we regard $X^{10}$ ({\sl not} $X^{y}$) as the 11th direction,
we should take T-dual along the $X^9$-direction to transform type IIA
superstring theory to type IIB superstring theory.
Then we can rewrite the background fields in terms of those of the
type IIB supergravity as follows (cf. Appendix \ref{S:R}),
\begin{eqnarray}
 \tg_{\mu\nu}
  &=&\sqrt{(\hp+\hq l)^2+e^{-2\varphi}\hq^2}\Bigl(\jmath_{\mu\nu}
    -\frac{\jmath_{9\mu}\jmath_{9\nu}
	-\Bpq_{9\mu}\Bpq_{9\nu}}{\jmath_{99}}\Bigr)\,,\\
 \tg_{\mu y}&=&
	\frac{\Bpq_{9\mu}}{\jmath_{99}}\,,\\
 \tg_{yy}&=&
    \frac{1}{\jmath_{99}\sqrt{(\hp+\hq l)^2+ e^{-2\varphi}\hq^2}}\,,\\
 \hq A_{\mu\nu9}+\hp A_{\mu\nu10}&=&
    \sqrt{(\hp+\hq l)^2+ e^{-2\varphi}\hq^2}\,\Bigl(\Bpq_{\mu\nu}
	+\frac{2\Bpq_{9[\mu}\jmath_{\nu]9}}{\jmath_{99}}\Bigr)\,,\\
  A_{\mu910}&=&-\frac{\jmath_{9\mu}}{\jmath_{99}}\,,
\end{eqnarray}
where $\BNS_{\mu\nu}$ and $\BR_{\mu\nu}$ are the NSNS and RR
second-rank antisymmetric tensors, respectively, $\jmath_{\hmu\hnu}$
are the metric in type IIB supergravity, $l=\hG_{910}/\hG_{1010}=A_9$
and
\begin{equation}
  \Bpq_{\hmu\hnu}\equiv \frac{\hp\BNS_{\hmu\hnu}
    +\hq\BR_{\hmu\hnu}}{\sqrt{(\hp+\hq l)^2+ e^{-2\varphi}\hq^2}}\,.
\end{equation}
Then, plugging these equations into eq.(\ref{eq:ddrac}) we have
\begin{eqnarray}
 S_{ddr}&=&\frac{2\pi T}{2}\int\!d\tau\!\int_0^{2\pi}
    \hspace{-1ex}d\sigma\,C\sqrt{(\hp+\hq l)^2+e^{-2\varphi}\hq^2}\nn
    &&\times\Biggl[\sqrt{-\tilde\gamma}\,
      \tilde\gamma^{ij}\,(\p_iX^\mu\p_jX^\nu\,\jmath_{\mu\nu}
    +2\,\p_i X^\mu\p_j \tX^y\,\jmath_{9\mu}
	+\p_i\tX^y \p_j\tX^y\,\jmath_{99})\nn
    &&\hspace{3ex}+ \epn^{ij}\,\p_i X^\mu \p_j X^\nu\Bpq_{\mu\nu}
	+2\epn^{ij}\,\p_i\tX^y\p_jX^\mu\,\Bpq_{9\mu}\Biggr]
    \,.\label{eq:pqac}
\end{eqnarray}
Once we regard $X^{10}$ as the 11th direction, the type IIA string
tension $T_s$ is given by $2\pi L_1T/\sqrt{\hG_{1010}}$ \cite{Sch}
since the 11d metric $\hG_{MN}$ is converted to the type IIA metric
$g_{\hmu\hnu}$ by the relation
$\hG_{\hmu\hnu}=g_{\hmu\hnu}/\sqrt{\hG_{1010}}$.
Also, if we assume that $l$ and $\varphi$ are constant and hence
$e^\varphi=\gIIB$, we have
\begin{equation}
  2\pi T C\sqrt{(\hp+\hq l)^2+e^{-2\varphi}\hq^2}
    = w_1~T_s \sqrt{(p+q l)^2+e^{-2\varphi}q^2}\equiv w_1~T_{pq}\,,
\end{equation}
where $T_{pq}$ is the tension of a $(p,q)$-string in type IIB
superstring theory \cite{Sch}.
Actually, we see that both of the NSNS and RR antisymmetric tensors
have coupled to $X^{\hat\alpha}\equiv(X^\mu,\tX^y)$ in eq.(\ref{eq:pqac}),
which implies that the reduced action (\ref{eq:pqac}) is, in fact,
that of $(p,q)$-strings.
Note that $w_1$ is just the number of copies of the resulting
$(p,q)$-strings. If we allow $q$ to be zero and take
$(p,q,r,s)=(1,0,0,1)$, we have the fundamental strings in type IIB
superstring theory. On the other hand, $(p,q,r,s)=(0,1,1,0)$ leads to
the strings which couple minimally with the RR B-field, i.e., the
D-strings.

%%%%%%%%%%%%%%%%%%%%%%%%%%%%%%%%%%
\section{Summary and discussion}
%%%%%%%%%%%%%%%%%%%%%%%%%%%%%%%%%%
In this paper, we have studied the double dimensional reduction of the
wrapped supermembrane on $\R^9\times T^2$ and explicitly derived the
bosonic sector of the $(p,q)$-string action in eq.(\ref{eq:pqac}).
This indicates that the supermembrane actually includes a
$(p,q)$-string as an excitation mode or object.
The (1,0)-string (F-string) is, of course, an effective mode
in a weak coupling region $\gIIB\ll1$, while the (0,1)-string
(D-string) in a strong coupling region $\gIIB\gg1$ for $l=0$.
However, the valid region to treat the $(p,q)$-string perturbatively
is still obscure and is deserved to be investigated
further.\footnote{Of course, a BPS saturated classical solution of the
$(p,q)$-string action (\ref{eq:pqac}) is valid irrespective of the
value of the string coupling $\gIIB$.}

The procedure of the double dimensional reduction here should be
realized on the matrix-regularized wrapped supermembrane on
$\R^9\times T^2$ \cite{UY4}, which will be reported elsewhere
\cite{OUY}.

In this paper we have considered classically to approach the boundary
of vanishing cycles of the 2-torus with the wrapped supermembrane.
On the other hand, Refs.\cite{SY,UY} studied quantum mechanical
justification of the double dimensional reduction in Ref.\cite{DHIS}.
In those references, the Kaluza-Klein modes associated with the
$\rho$-coordinate were not removed classically, but they were
integrated in the path integral formulation of the wrapped
supermembrane theory.
Similar quantum mechanical investigation of the double dimensional
reduction in this paper deserves to be investigated.

%%%%%%%%%%%%%%%%%%%%%%%%%%%%%%%%%%%
\vspace{\baselineskip}

\noindent{\bf Acknowledgments:}
This work is supported in part by MEXT Grant-in-Aid for
the Scientific Research \#13135212 (S.U.).

\appendix
%S%%%%%%%%%%%%%%%%%%%%%%%%%%%%%%%%%%%%%%%%%%%%%%%%%%%%%%%%%%%%%%
\section{Notation}
The spacetime indices:
\begin{eqnarray}
  M,N,P &=& 0,1,\dots,8,9,10\,, \\
  U,V &=& 0,1,\dots,8,y,z\,, \\
  \hmu,\hnu &=& 0,1,\dots,8,9\,, \\
  \mu,\nu &=& 0,1,\dots,8\,.
\end{eqnarray}
The worldvolume and worldsheet indices:
\begin{eqnarray}
  \alpha ,\beta  &=& 0,1,2\,, \\
  i,j &=& 0,1\,.
\end{eqnarray}
The target-space metrics:
\begin{eqnarray}
  G&=& \textrm{11d target-space metric} \,,\\
  \tG&=& \textrm{11d rotated target-space metric} \,,\\
  g&=& \textrm{10d IIA target-space metric}\,,\\
  \tg &=& \textrm{10d IIA rotated target-space metric} \,,\\
  \jmath &=& \textrm{10d IIB target-space metric} \,.
\end{eqnarray}
The worldvolume and worldsheet metrics:
\begin{eqnarray}
\hat{\gamma} &=& \textrm{membrane worldvolume metric}\,,\\
\gamma &=& \textrm{string worldsheet metric}\,.
\end{eqnarray}
(Anti-)symmetrization r.w.t.\ indices:
\begin{eqnarray}
   A_{[\mu} B_{\nu]} &=& \frac{1}{2}\left(A_{\mu} B_{\nu}
	- A_{\nu} B_{\mu}\right)\,,\\
   A_{[\mu} B_{\nu} C_{\rho]} &=& \frac{1}{3!}
	(A_{\mu} B_{\nu}C_{\rho}
	+A_{\nu} B_{\rho}C_{\mu}+A_{\rho} B_{\mu}C_{\nu}\nn
    &&\quad - A_{\mu}B_{\rho}C_{\nu}
	-A_{\rho} B_{\nu}C_{\mu}-A_{\nu} B_{\mu}C_{\rho})\,,\\
   A_{[\mu} B_{|\nu|} C_{\rho]} &=& \frac{1}{2}
	(A_{\mu} B_{\nu}C_{\rho}-A_{\rho} B_{\nu}C_{\mu})\,,\\
   A_{\{\mu} B_{\nu\}} &=& \frac{1}{2}\left(A_{\mu} B_{\nu}
	+ A_{\nu} B_{\mu}\right)\,,\quad\mbox{etc.}
\end{eqnarray}

%%%%%%%%%%%%%%%%%%%%%%%%%%%%%%%%%%%%%%%%%%%%%%%%%%
\section{11d  vs. 10d background fields}\label{S:R}
%%%%%%%%%%%%%%%%%%%%%%%%%%%%%%%%%%%%%%%%%%%%%%%%%%
The 11-dimensional metric can be written by
\begin{eqnarray}
 \hG_{MN} &=& \left(\begin{array}{@{\,}cc@{\,}}
	\frac{1}{\sqrt{\hG_{1010}}}\,\ga_{\hmu\hnu}
	+\frac{1}{\hG_{1010}}\hG_{\hmu10}\hG_{\hnu10}&
		 \hG_{\hmu10} \\[10pt]
	\hG_{\hnu10} & \hG_{1010} \end{array}\right)\nn
   &\equiv& e^{-\frac{2}{3}\phi}
    \left(\begin{array}{@{\,}cc@{\,}}
	\ga_{\hmu\hnu}+e^{2\phi}A_{\hmu}A_{\hnu}&
		 e^{2\phi}  A_{\hmu} \\[10pt]
	e^{2\phi}  A_{\hnu} & e^{2\phi} \end{array}\right)\nn
  &=& \left(\begin{array}{@{\,}ccc@{\,}}
	\frac{1}{\sqrt{\hG_{1010}}}\,\ga_{\mu\nu}
	+\frac{1}{\hG_{1010}}\hG_{\mu10}\hG_{\nu10}&
	\frac{1}{\sqrt{\hG_{1010}}}\,\ga_{\mu9}
	+\frac{1}{\hG_{1010}}\hG_{\mu10}\hG_{910}& \hG_{\mu10}\\[10pt]
	\frac{1}{\sqrt{\hG_{1010}}}\,\ga_{9\nu}
	+\frac{1}{\hG_{1010}}\hG_{910}\hG_{\nu10}&
	\frac{1}{\sqrt{\hG_{1010}}}\,\ga_{99}
	+\frac{1}{\hG_{1010}}\hG_{910}\hG_{910}& \hG_{910} \\[10pt]
	\hG_{\nu10}&\hG_{910} & \hG_{1010}
			\end{array}\right)\,,
\end{eqnarray}
and the third-rank antisymmetric tensor $A_{MNP}$ is decomposed as
\begin{eqnarray}
 A_{MNP} &=& (A_{\mu\nu\rho}, A_{\mu\nu10},A_{\mu\nu9},A_{\mu910})\nn
     &=& (C_{\mu\nu\rho}, B_{\mu\nu},C_{\mu\nu9},
	B_{\mu9})\,.
\end{eqnarray}
Those fields are related to those of IIB as
\begin{eqnarray}
  \ga_{\mu\nu}&=& \jmath_{\mu\nu}
    -\frac{\jmath_{9\mu}\jmath_{9\nu}
	-\BNS_{9\mu}\BNS_{9\nu}}{\jmath_{99}}\,,\\
  \ga_{9\mu}&=& \frac{\BNS_{9\mu}}{\jmath_{99}}\,,\\
  \ga_{99}&=&\frac{1}{\jmath_{99}}\,,\\
  C_{\hmu\nu9}&=&\BR_{\mu\nu}
	+\frac{2\BR_{9[\mu}\jmath_{\nu]9}}{\jmath_{99}}\,,\\
  C_{\mu\nu\rho}&=&D_{9\mu\nu\rho}
	+\frac{3}{2}\,\epn^{ij}B^{(i)}_{9[\mu}\,B^{(j)}_{\nu\rho]}
	+\frac{3}{2}\,\epn^{ij}\frac{B^{(i)}_{9[\mu}\,
	B^{(j)}_{\nu|9|}\jmath_{\rho]9}}{\jmath_{99}}\,,\\
  B_{\mu\nu}&=& \BNS_{\mu\nu}+\frac{\BNS_{9\mu}\jmath_{\nu9}
	-\BNS_{9\nu}\jmath_{\mu9}}{\jmath_{99}}\,,\\
  B_{9\mu}&=& \frac{\jmath_{9\mu}}{\jmath_{99}}\,,\\
  A_\mu &=& -\BR_{9\mu} + l \BNS_{9\mu}\,,\\
  A_9 &=& l\,,\\
  \phi&=&\varphi -\frac{1}{2}\ln \jmath_{99}\,.
\end{eqnarray}

On the other hand, the 9-10 rotated metric is given by
($U,V=0,1,\cdots,8,y,z$)
\begin{eqnarray}
 \tG_{UV} &=& \hG_{MN}\,\frac{\p X^M}{\p X^U}\,
	\frac{\p X^N}{\p X^V}\nn
  &=& \left(\begin{array}{@{\,}ccc@{\,}}
	\frac{1}{\sqrt{\tG_{zz}}}\,\tg_{\mu\nu}
	+\frac{1}{\tG_{zz}}\tG_{\mu z}\tG_{\nu z}&
	\frac{1}{\sqrt{\tG_{zz}}}\,\tg_{\mu y}
	+\frac{1}{\tG_{zz}}\tG_{\mu z}\tG_{yz}& \tG_{\mu z} \\[10pt]
	\frac{1}{\sqrt{\tG_{zz}}}\,\tg_{y\nu}
	+\frac{1}{\tG_{zz}}\tG_{yz}\tG_{\nu z}&
	\frac{1}{\sqrt{\tG_{zz}}}\,\tg_{yy}
	+\frac{1}{\tG_{zz}}\tG_{yz}\tG_{yz}& \tG_{yz} \\[10pt]
	\tG_{\nu z}&\tG_{yz} & \tG_{zz}
			\end{array}\right)\,.
\end{eqnarray}
Then
\begin{eqnarray}
 \tG_{zz}&=&\hq^2\,\hG_{99}+2\hp\hq\,\hG_{910}
		+\hp^2\,\hG_{1010}\nn
  &=&\hq^2\left(\frac{\ga_{99}}{\sqrt{\hG_{1010}}}
    +\hG_{1010}A_9^2\right) +2\hp\hq\hG_{1010}A_9+\hp^2\hG_{1010}\nn
  &=&\hG_{1010}(\hp+\hq A_9)^2
	+\hq^2\frac{\ga_{99}}{\sqrt{\hG_{1010}}}
  =e^{4\varphi/3}\jmath_{99}^{-2/3}\left\{(\hp+\hq l)^2
	+e^{-2\varphi}\hq^2\right\}\,,\label{eq:2zz}\\
 \tG_{yy}&=&\hp^2\,\hG_{99}-2\hp\hq\,\hG_{910}
		+\hq^2\,\hG_{1010}\nn
  &=&\hp^2\left(\frac{\ga_{99}}{\sqrt{\hG_{1010}}}
	+\hG_{1010} A_9^2\right)
	-2\hp\hq\hG_{1010}A_9+\hq^2\hG_{1010}\nn
  &=&\hG_{1010}(\hq-\hp A_9)^2
	+\hp^2\frac{\ga_{99}}{\sqrt{\hG_{1010}}}
  =e^{4\varphi/3}\jmath_{99}^{-2/3}\left\{(\hq-\hp l)^2
	+\hp^2e^{-2\varphi}\right\}\,,\label{eq:2yy}\\
 \tG_{yz}&=&\hp\hq\,\hG_{99}+(\hp^2-\hq^2)\,\hG_{910}
		-\hp\hq\,\hG_{1010}\nn
  &=&\hp\hq\left(\frac{\ga_{99}}{\sqrt{\hG_{1010}}}
	+\hG_{1010} A_9^2\right)
	+(\hp^2-\hq^2)\hG_{1010}A_9-\hp\hq\hG_{1010}\nn
  &=&\hG_{1010}(\hp A_9-\hq)(\hq A_9+\hp)
	+\hp\hq\frac{\ga_{99}}{\sqrt{\hG_{1010}}}\nn
  &=&e^{4\varphi/3}\jmath_{99}^{-2/3}\left\{(\hp l-\hq)(\hq l+\hp)
	+\hp\hq e^{-2\varphi}\right\}\,,\label{eq:2yz}
\end{eqnarray}
Furthermore,
\begin{equation}
 \tG_{\mu y}= \hp\,\hG_{\mu9}-\hq\,\hG_{\mu10}
	 = \frac{1}{\sqrt{\tG_{zz}}}\,\tg_{\mu y}
	+\frac{1}{\tG_{zz}}\tG_{\mu z}\tG_{yz}\,,
\end{equation}
and hence
\begin{eqnarray}
 \tg_{\mu y}&=&-\frac{1}{\sqrt{\tG_{zz}}}\left(
	\tG_{\mu y}\tG_{zz}-\tG_{\mu z}\tG_{yz}\right)\nn
 &=&-\frac{1}{\sqrt{\tG_{zz}}}\Biggl\{
    (\hp\,\hG_{\mu9}-\hq\,\hG_{\mu10})
	(\hq^2\,\hG_{99}+2\hp\hq\,\hG_{910}
		+\hp^2\,\hG_{1010})\nn
 &&\hspace{10ex}-(\hq\,\hG_{\mu9}+\hp\,\hG_{\mu10})
	(\hp\hq\,\hG_{99}+(\hp^2-\hq^2)\,\hG_{910}
	-\hp\hq\,\hG_{1010})\Biggr\}\nn
 &=&\frac{\hG_{\mu9}\,(\hp\,\hG_{1010}+\hq\,\hG_{109})
    -\hG_{\mu10}\,(\hq\,\hG_{99} +\hp\,\hG_{910})}{\sqrt{\tG_{zz}}}\nn
 &=&\sqrt{\frac{\hG_{1010}}{\tG_{zz}}}\,\left\{(\hp+\hq A_9) \ga_{\mu9}
	-\hq \ga_{99}A_\mu\right\}\nn
 &=&\sqrt{\frac{\hG_{1010}}{\tG_{zz}}}\,\left\{(\hp+\hq l)
	 \frac{\BNS_{9\mu}}{\jmath_{99}}
	-\hq\frac{-\BR_{9\mu}+l\BNS_{9\mu}}{\jmath_{99}}\right\}
 =\frac{\hp\,\BNS_{9\mu}+\hq\,\BR_{9\mu}}{%
	\jmath_{99}\sqrt{(\hp+\hq l)^2+e^{-2\varphi}\hq^2}}\,.~
\end{eqnarray}
We shall calculate $\tg_{\mu\nu},\tg_{yy}$ as follows.
The equation,
\begin{equation}
  \tG_{yy}= \frac{1}{\sqrt{\tG_{zz}}}\,\tg_{yy}
	+\frac{1}{\tG_{zz}}\tG_{yz}\tG_{yz}\,,
\end{equation}
leads to
\begin{eqnarray}
 \tg_{yy}&=&\frac{1}{\sqrt{\tG_{zz}}}\left(
	\tG_{yy}\tG_{zz}-\tG_{yz}\tG_{yz}\right)\nn
 &=&\frac{1}{\sqrt{\tG_{zz}}}\Biggl[
    \Bigl\{\hG_{1010}(\hq-\hp A_9)^2
	+\hp^2\frac{\ga_{99}}{\sqrt{\hG_{1010}}}\Bigr\}
    \Bigl\{\hG_{1010}(\hp+\hq A_9)^2
	+\hq^2\frac{\ga_{99}}{\sqrt{\hG_{1010}}}\Bigr\}\nn
 &&\hspace{10ex}-\Bigl\{\hG_{1010}(\hp A_9-\hq)(\hq A_9+\hp)
	+\hp\hq\frac{\ga_{99}}{\sqrt{\hG_{1010}}}\Bigr\}^2\Biggr]\nn
 &=&\sqrt{\frac{\hG_{1010}}{\tG_{zz}}}\,\ga_{99}\Biggl\{
	\hp^2(\hp+\hq A_9)^2+\hq^2(\hq-\hp A_9)^2
	-2\hp\hq(\hp A_9-\hq)(\hq A_9+\hp) \Biggr\}\nn
 &=&\sqrt{\frac{\hG_{1010}}{\tG_{zz}}}\,\ga_{99}
    =\frac{1}{\jmath_{99}\sqrt{(\hp+\hq l)^2
		+ e^{-2\varphi}\hq^2}}\,.
\end{eqnarray}
Similarly
\begin{equation}
 \hG_{\mu\nu}= \frac{1}{\sqrt{\hG_{1010}}}\,\ga_{\mu\nu}
	+\frac{1}{\hG_{1010}}\hG_{\mu10}\hG_{\nu10}
	 = \frac{1}{\sqrt{\tG_{zz}}}\,\tg_{\mu\nu}
	+\frac{1}{\tG_{zz}}\tG_{\mu z}\tG_{\nu z}\,,
\end{equation}
leads to
\begin{eqnarray}
 \tg_{\mu\nu}&=&\frac{1}{\sqrt{\tG_{zz}}}\left(
	\hG_{\mu\nu}\tG_{zz}-\tG_{\mu z}\tG_{\nu z}\right)\nn
 &=&\frac{1}{\sqrt{\tG_{zz}}}\Biggl\{
	\Bigl(\frac{1}{\sqrt{\hG_{1010}}}\,\ga_{\mu\nu}
	+\frac{1}{\hG_{1010}}\hG_{\mu10}\hG_{\nu10}\Bigr)\tG_{zz}
	-(\hq\hG_{\mu9}+\hp\hG_{\mu10})(\hq\hG_{\nu9}
	+\hp\hG_{\nu10})\Biggr\}\nn
 &=&\sqrt{\frac{\tG_{zz}}{\hG_{1010}}}\,\ga_{\mu\nu}
	+\frac{\tG_{zz}\hG_{\mu10}\hG_{\nu10}
	-\hG_{1010}(\hq\hG_{\mu9}+\hp\hG_{\mu10})(\hq\hG_{\nu9}
	+\hp\hG_{\nu10})}{\hG_{1010}\sqrt{\tG_{zz}}}\nn
 &=&\sqrt{\frac{\tG_{zz}}{\hG_{1010}}}\,\ga_{\mu\nu}
	+\hq^2\,\sqrt{\frac{\hG_{1010}}{\tG_{zz}}}\,\Biggl\{
	\ga_{99}A_\mu A_\nu
	-\frac{\ga_{\mu9}\ga_{\nu9}}{(\hG_{1010})^{3/2}}
	-A_9(A_\mu \ga_{\nu9}+ A_\nu \ga_{\mu9})\Biggr\}\nn
 &&\qquad-\,\hp\hq\,\sqrt{\frac{\hG_{1010}}{\tG_{zz}}}\,
	(A_\mu \ga_{\nu9}+A_\nu \ga_{\mu9})\nn
 &=&\sqrt{\frac{\tG_{zz}}{\hG_{1010}}}\,\ga_{\mu\nu}
    +\hq^2\,\sqrt{\frac{\hG_{1010}}{\tG_{zz}}}\,
    \Biggl[\frac{(-\BR_{9\mu}+l\BNS_{9\mu})(-\BR_{9\nu}
	+l\BNS_{9\nu})}{\jmath_{99}}
    -\frac{\BNS_{9\mu}\BNS_{9\nu}}{e^{2\varphi}\jmath_{99}}\nn
 &&\hspace{20ex}-l\,\frac{(-\BR_{9\mu}+l\BNS_{9\mu})\BNS_{9\nu}
    +(-\BR_{9\nu}+l\BNS_{9\nu})\BNS_{9\mu}}{\jmath_{99}}\Biggr]\nn
 &&\qquad-\,\hp\hq\,\sqrt{\frac{\hG_{1010}}{\tG_{zz}}}\,
	\frac{(-\BR_{9\mu}+l\BNS_{9\mu})\BNS_{9\nu}
    +(-\BR_{9\nu}+l\BNS_{9\nu})\BNS_{9\mu}}{\jmath_{99}}\nn
 &=&\sqrt{\frac{\tG_{zz}}{\hG_{1010}}}\,\ga_{\mu\nu}
    +\frac{1}{\jmath_{99}}\,\sqrt{\frac{\hG_{1010}}{\tG_{zz}}}\,
    \Biggl[(\hp\BNS_{9\mu}+\hq\BR_{9\mu})
	(\hp\BNS_{9\nu}+\hq\BR_{9\nu})\nn
 &&\hspace{30ex}-\{(\hp+\hq l)^2 +\hq^2\,
	e^{-2\varphi}\}\BNS_{9\mu}\BNS_{9\nu}\Biggr]\nn
 &=&\sqrt{(\hp+\hq l)^2+e^{-2\varphi}\hq^2}\,\left(\ga_{\mu\nu}
	-\frac{\BNS_{9\mu}\BNS_{9\nu}}{\jmath_{99}}\right)
 +\,\frac{(\hp\BNS_{9\mu}+\hq\BR_{9\mu})(\hp\BNS_{9\nu}
    +\hq\BR_{9\nu})}{\jmath_{99}\,\sqrt{(\hp+\hq l)^2
	+e^{-2\varphi}\hq^2}}\nn
 &=&\sqrt{(\hp+\hq l)^2+e^{-2\varphi}\hq^2}\,\left(\jmath_{\mu\nu}
	-\frac{\jmath_{9\mu}\jmath_{9\nu}}{\jmath_{99}}
    +\,\frac{(\hp\BNS_{9\mu}+\hq\BR_{9\mu})(\hp\BNS_{9\nu}
    +\hq\BR_{9\nu})}{\jmath_{99}\,\{(\hp+\hq l)^2
	+e^{-2\varphi}\hq^2\}}\right) \,.
\end{eqnarray}
Note that
\begin{equation}
 \sqrt{\frac{\tG_{zz}}{\hG_{1010}}}=
	\sqrt{(\hp+\hq l)^2+e^{-2\varphi}\hq^2}\,.
\end{equation}

%%%%%%%%%%%%%%%%%%%%%%%

\end{document}